\documentclass[prb,showpacs,preprintnumbers,preprint,superscriptaddress]{revtex4}

\usepackage{graphicx}
\usepackage{dcolumn}
\usepackage{bm}
\usepackage{amsmath}
\usepackage{amsfonts}
\usepackage{amssymb}
\usepackage[latin1]{inputenc}

\begin{document}
%%%%%%%%%%%%%%%%%%%%%%%%%%%%%%%%%%%%%%%%%%%%%%%%%%%%%%%%%%%%%%%%%
%%%%%%%%%%%%%%%%%%%%%%%%%%%%%%%%%%%%%%%%%%%%%%%%%%%%%%%%%%%%%%%%%

\title{Efficient treatment of stacked metasurfaces for optimizing and enhancing the range of accessible optical functionalities}

\author{C. Menzel}
\affiliation{Institute of Applied Physics, Abbe Center of Photonics,\\ Friedrich-Schiller-Universit\"at Jena, Albert-Einstein-Str. 15, 07745 Jena, Germany}
\author{J. Sperrhake}
\affiliation{Institute of Applied Physics, Abbe Center of Photonics,\\ Friedrich-Schiller-Universit\"at Jena, Albert-Einstein-Str. 15, 07745 Jena, Germany}
\author{T. Pertsch}
\affiliation{Institute of Applied Physics, Abbe Center of Photonics,\\ Friedrich-Schiller-Universit\"at Jena, Albert-Einstein-Str. 15, 07745 Jena, Germany}

\begin{abstract}
We present, discuss and validate an adapted S-matrix formalism for an efficient, simplified treatment of stacked homogeneous periodically structured metasurfaces operated under normally incident plane wave excitation. The proposed formalism can be applied to any material system, arbitrarily shaped metaatoms, at any frequency and with arbitrary subwavelength periods. Circumventing the introduction of any kind of effective parameters we directly use the S-parameters of the individual metasurfaces to calculate the response of an arbitrary stack. In fact, the S-parameters are the complex parameters of choice fully characterizing the homogeneous metasurfaces, in particular with respect to its polarization manipulating properties. Just as effective material parameters like the permittivity and the permeability or wave parameters like the propagation constant and the impedance, the stacking based upon S-matrices can be applied as long as the individual layers are decoupled with respect to their near-fields.
%In terms of periodically structured surfaces and hence Bloch modes, this requires the individual layers to fulfill the so-called fundamental mode approximation with respect to plane wave excitation.
This requirement eventually sets the limits for using the optical properties of the individual layers to calculate the response of the stacked system - this being the conceptual aim for any homogeneous metasurface or metamaterial layer and therefore the essence of what is eventually possible with homogeneous metasurfaces. As simple and appealing this approach is, as powerful it is as well: Combining structured metasurface with each other as well as with isotropic, anisotropic or chiral homogeneous layers is possible by simple semi-analytical S-matrix multiplication. Hence, complex stacks and resonators can be set up, accurately treated and optimized with respect to their dispersive polarization sensitive optical functionality without the need for further rigorous full-wave simulations.
\end{abstract}

%  42.25.Bs     - Wave propagation transmission and absorption
%  78.20.-e 	- optical properties of bulk and thin films
%  78.67.Pt 	- Multilayers; superlattices; photonic structures; metamaterials
%  81.05.Xj 	- Metamaterials for chiral, bianisotropic and other complex media

\maketitle

\section{Introduction}
Metamaterials, i.e. artificial sub-wavelength structured materials\cite{Shivola2007,Shamonina2007}, attracted a great deal of interest on all wavelength scales ranging from mm-waves to optics for already more than one and a half decades\cite{Pendry1998,Shelby2001,Smith2004,Soukoulis2007Science}. Where the early focus was on the realization of artificial, usually periodically structured materials with tailored material properties for full control of propagation, dispersion and polarization, a new class, most often called metasurfaces, emerged taking control over diffraction as well\cite{Ni2012,Yu2011,Yu2014,AluTopicalReview}. Common to both classes is their composition of metaatoms each of them being sub-wavelength in its lateral dimensions. Here, a single layer of metaatoms will be called metasurface irrespective of the shape and composition of its individual metaatoms. To further distinguish between both classes we will call metasurfaces comprised of identical metaatoms with subwavelength inter-particle distances \textit{homogeneous} metasurfaces, which are also known as frequency-selective surfaces (FSS)\cite{Mitra1988,Sievenpiper1999,Munk2000,Debus2007}. Their far-field response is fully contained in a zeroth diffraction order in transmission and reflection \footnote{Aperiodically or amorphously arranged identical metaatoms, designed for controlling the zeroth diffraction orders and, hence, neglecting scattering losses are called homogeneous metasurfaces as well.}. Metamaterials are then understood as stacked identical homogeneous metasurfaces. In contrast, an \textit{inhomogeneous} metasurface with gradually or abruptly varying arrangements of metaatoms across the surface allows the control of a larger number of diffraction orders and can be understood as a hologram\cite{Larouche2012,Levy2007,Ni2013,Sun2012,Walther2011,Walther2012,RepProgPhysCapasso} in its most general sense. In the present manuscript we will deal with the stacking of homogeneous metasurfaces (MS) only.

Exploiting a stacking of metasurfaces to enhance the range of accessible optical functionalities is widely used e.g. for tailoring dispersion\cite{Ranjan2015}, diffraction\cite{PaulAdvMat} and in particular for controlling the polarization state of light\cite{Svirko2001,Stereometamaterials,DeckerOL2009,DeckerOL2010,Ozbay2010,MutluJOpt2013}. However, just a limited number of publications explicitly dealt with the stacking of decoupled homogeneous metasurfaces\cite{AluStack1,AluStack2,GrbicStack}, where our approach - based on the S-matrix of the individual MS - is fundamentally different.

Originally, the individual homogeneous metasurfaces,
%
%created to realize new materials with electromagnetic properties not available in nature,
%
ought to be described by universal material properties reducing the generally complicated electromagnetic response of periodically structured surfaces to a few parameters only. Unfortunately, it turned out that these parameters depend on the embedding of the MS\cite{AlbooyehSubstrate} and might change upon stacking of identical MSs. The reason for this lies in the near-field coupling of the MS with its surrounding\cite{Simovski2007,Rockstuhl2007PRB,Tserkezis2008,Zhou2009,MenzelSCIC2010,Andryieuski2010,Paul2011}. Furthermore, for MSs comprised of low-symmetry metaatoms being ideally described by bianisotropic constitutive relations\cite{SerdyukovBook,ChenBianisotropicRetrieval2005,Silveirinha2007}, the retrieval of effective material parameters becomes cumbersome. Eventually, most MSs operating in resonant regime exhibit a strongly non-local response disqualifying the use of local effective material parameters\cite{Simovski2007b,SimovskiReview,Cabuz2008,MenzelValidity,Tserkezis2010,Alu2011}, which do not depend on the wavevector or angle of incidence. In particular in the optical domain just a single publication is known where a local description of an artificial magnetic response is validated\cite{MenzelOL2012}.

\begin{figure*}[t]
\begin{center}
\includegraphics[width=150mm,angle=0] {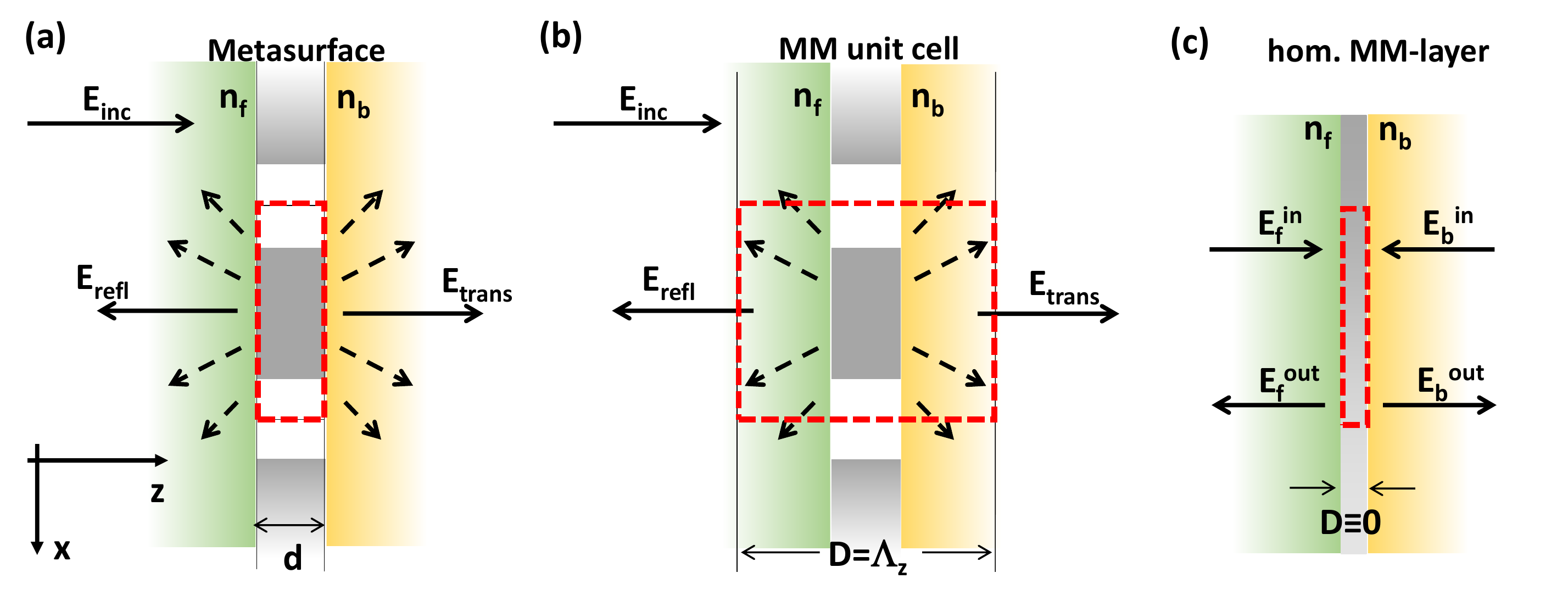}
\caption{Schematic of the geometry under consideration. The figures show a $xz$-cut of the structured surface periodic in $x$ and $y$-direction ($\Lambda_x,\Lambda_y$). The surface is embedded in halfspaces characterized by refractive indices $n_\mathrm{f}$ and $n_\mathrm{b}$ in front and back of the surface. The solid arrows indicate zeroth diffraction orders. The dashed arrows indicate evanescent diffraction orders. The red dashed line contains the periodic unit cell. (a) Metasurface/grating with physical thickness $d$ of the structured surface. (b) The same MS as in (a) with additional spacing layers defining the new MM unit cell with period $D=\Lambda_z$ in z-direction. At the z-boundary of the new unit cell the evanescent diffraction orders are sufficiently decayed such that the field is plane wave like. (c) The homogeneous MM unit cell described in (b) is replaced by an effective homogeneous MM-layer with virtual thickness $D\equiv 0$. Due to the translational invariance along the $x$ and $y$-direction the definition of the unit cell is arbitrary and indicated here just to anticipate the transition from (b) to (c). Such layers are the building blocks of the considered stacked MSs.}
\label{FIG_IntroGeneral3}
\end{center}
\end{figure*}
Once the electromagnetic properties cannot be reduced to local material parameters, we can remain on the level of generally wave vector dependent dispersion relations for the propagation constant $k_\mathrm{prop}(\mathbf{k}_\mathrm{t})$ and Bloch impedances\cite{Simovski2007,Paul2011} $Z(\mathbf{k}_\mathrm{t})$, which are available e.g. via the S-parameter retrieval\cite{Paul2011,MenzelValidity,Smith2002,Menzel2008PRB} or similar methods\cite{WavePropAndrei,PaulHalfSpace,BlochAnalysisAndrei}. Under certain circumstances, namely the validity of the fundamental Bloch mode approximation (FMA)\cite{Paul2011}, the reduction to thickness independent $k$ of the fundamental Bloch mode and its Bloch impedance $Z$ is possible and undoubtedly useful for stacked systems of identical MSs. Here, the effective parameters $k$ and $Z$ are in fact independent of the number of layers\cite{Simovski2007}. The validity of the FMA is of major importance for the stacking in general: Only for MSs fulfilling the FMA the far-field response of the stacked system can be calculated rigorously from the far-field response of the individual MSs.

However, if non-identical MSs ought to be stacked for optimizing a specific optical functionality and, hence, the overall far-field response (transmission and reflection), the treatment of the individual MSs by effective parameters is not meaningful. It suffices to remain on the equivalent level of zeroth order transmission ($t$) and reflection ($r$) for describing the individual MS, thereby circumventing any kind of retrieval procedure. We just have to combine the $r$ and $t$ of the individual MSs appropriately to get $r$ and $t$ of the stacked system - this being the aim of the present manuscript\footnote{This formalism was first presented at the META'15 conference.}. In fact, it captures the essence of what is eventually possible with homogeneous MSs and what their conceptual design guideline was: the reduction of the complex response of the individual MS to a few essential parameters and use of these parameters for the rigorous determination of the properties of an arbitrarily stacked MM system.

The essential parameters describing the MSs are their complex 4x4 S-matrices\cite{Li1996}, comprised of the forward and backward reflection and transmission coefficients $r_{ij}$ and $t_{ij}$. They can be determined either by rigorous simulations, on analytical grounds or by experimental characterization even in the optical domain \cite{Helgert2011,Pshenay2014}. The analytical calculation is of particular importance: As complexity of the response in particular with respect to polarization can be achieved by stacking, the individual MSs can be realized as simple planar MSs, that can be efficiently modelled as arrays of coupled electric and magnetic dipoles\cite{Tretyakov2003,Belov2005,Sersic2009,Simovski2011}. By additional use of the stacking-algorithm presented here, the overall response of the stacked system can be modelled analytically and efficiently optimized. Furthermore, restricting to planar MSs is advantageous for systems operating in the NIR and VIS domain significantly simplifying their fabrication compared to MSs composed of complex shaped 3D metaatoms \cite{Helgert2011,Schaferling2012,Kaschke2015} and obviating the subtle issue of lateral alignment of subsequent layers\cite{AluStack2}.

The stacking algorithm\footnote{The proposed stacking formalism is basically a modified S-matrix formalism for stacking of homogeneous media.} as presented can be applied to any kind of subwavelength structured homogeneous MS with arbitrarily shaped metaatoms irrespective of the material system and the wavelength. The different MSs can have similar or different as well as incommensurable periods, which cannot be treated on rigorous grounds by numerical simulations. Within the stack common optical materials like isotropic or chiral materials and anisotropic crystals can be used as well.

The remainder of the manuscript is outlined as follows: In Sec.~2 we define the system under consideration and discuss the representation of the periodically structured system with respect to the reduced S-matrix. In Sec.~3 we present the formulas necessary for the stacking, provide an estimate for the necessary critical embedding thickness validating the FMA and discuss symmetry operations on S-matrices. In Sec.~4 we discuss some prototypical examples by comparing the rigorous and approximated solution based on the reduced S-matrix. We conclude the manuscript in Sec.~5.
\section{Introducing the S-matrix}
We assume systems that are periodic in $x$ and $y$-direction with periods $\Lambda_x$ and $\Lambda_y$ and plane wave propagation along the $z$-direction with wave number $k$ and frequency $\omega$, hence an incident electric field of the form
\begin{equation}
\mathbf{E}_\mathrm{inc}=\left(E_x\vec{e}_x+E_y\vec{e}_y\right)e^{i(kz-\omega t)}.
\end{equation}
The periodically structured MS [Fig.~\ref{FIG_IntroGeneral3}(a)] acts as a sub-wavelength grating, where in general an infinite number of diffraction orders, i.e. plane wave expansion coefficients, of the overall field on both sides have to be taken into account for a rigorous description including the near field\cite{Paul2011}. However, for a subwavelength grating with a free space wavelength $\lambda>\max[n_\mathrm{f},n_\mathrm{b}]\cdot\max[\Lambda_x,\Lambda_y]$ all higher diffraction orders are evanescent for normal incidence. Only the zeroth diffraction order in reflection and transmission are non-evanescent (see Fig.~\ref{FIG_IntroGeneral3}) contributing to the far-field response.
The response of such a system schematically shown in Fig.~\ref{FIG_IntroGeneral3}(a) strongly depends on the embedding and any other MS placed closely in front or back of the first one effects the response due to near-field coupling mediated by the evanescent fields between both\cite{Paul2011}. As is well known, the near-field coupling disqualifies any effective medium approach and the response of the combined or stacked system has to be treated rigorously taking into account all evanescent diffraction orders as well. To obviate the near-field coupling we have to assure a minimum distance between different MSs or the MS and any interface to homogeneous layers introducing a new thickness $D=\Lambda_z$ which defines the unit cell in $z$-direction [see Fig.~\ref{FIG_IntroGeneral3}(b)].

%
% too much redundancy?
%
In terms of Bloch modes, the newly created MM unit cell satisfies the fundamental Bloch mode approximation (FMA) with respect to plane wave coupling\cite{Paul2011}. The fundamental Bloch mode of the periodic system is plane wave like at the boundaries and the system is fully described by its zeroth order transmission and reflection coefficients for plane wave excitation. In fact, such a system can be described by effective wave parameters which are the propagation constant of the fundamental mode and its Bloch impedance \cite{Simovski2007,Paul2011,BlochAnalysisAndrei}. However, for low-symmetry MS the Bloch impedance becomes tensorial and two propagation constants need to be considered for reciprocal systems. To avoid the issue of introducing and retrieving these effective wave parameters, we remain on the level of complex reflection and transmission coefficients, which become $2\mathrm{x}2$ matrices for low-symmetry MM and, hence, 4x4 matrices taking into account both propagation directions.

A single MM layer that fulfills the FMA is called homogeneous MM and can be replaced conceptually by a single complex layer with virtual thickness $D=0$ as shown in Fig.~\ref{FIG_IntroGeneral3}(c), i.e. a true MS. Its response upon normally incident plane wave excitation is fully characterized by the S-matrix defined below.

The plane wave field in front (f) and back (b) of such a system can be written as
\begin{equation}
\mathbf{E}^\mathrm{f}(z<0)=\left[\mathbf{E}^\mathrm{f}_\mathrm{in}e^{ikz}+\mathbf{E}^\mathrm{f}_\mathrm{out}e^{-ikz}\right]e^{-i\omega t}\end{equation}
\begin{equation}
\mathbf{E}^\mathrm{b}(z>0)=\left[\mathbf{E}^\mathrm{b}_\mathrm{out}e^{ikz}+\mathbf{E}^\mathrm{b}_\mathrm{in}e^{-ikz}\right]e^{-i\omega t}\end{equation}
The S-matrix describing the plane-wave response of the system connects the incoming and outgoing complex two-component field vectors $\mathbf{E}=(E_x,E_y)^T$
\begin{equation}
\begin{pmatrix} \mathbf{E}^\mathrm{b}_\mathrm{out} \\ \mathbf{E}^\mathrm{f}_\mathrm{out} \end{pmatrix}=
\mathbf{S}\begin{pmatrix} \mathbf{E}^\mathrm{f}_\mathrm{in} \\ \mathbf{E}^\mathrm{b}_\mathrm{in}\end{pmatrix} =
\begin{pmatrix} \hat{S}_{11} & \hat{S}_{12} \\ \hat{S}_{21} & \hat{S}_{22} \end{pmatrix}
\begin{pmatrix} \mathbf{E}^\mathrm{f}_\mathrm{in} \\ \mathbf{E}^\mathrm{b}_\mathrm{in} \end{pmatrix}
\end{equation}
For polarization insensitive samples, where no polarization rotation occurs, the sub-matrices $\hat{S}_\mathrm{ij}$ are scalars and directly give the complex transmission and reflection in forward ($t^\mathrm{f},r^\mathrm{f}$) and backward ($t^\mathrm{b},r^\mathrm{b}$) direction
\begin{equation}
\mathbf{S}=\begin{pmatrix} t^\mathrm{f} & r^\mathrm{b} \\ r^\mathrm{f} & f^\mathrm{b} \end{pmatrix}.
\end{equation}
For any sample affecting the polarization state in transmission or reflection the situation is more involved. The S-matrix in terms of transmission and reflection matrices indicated by capital letters is given as
\begin{equation}
\mathbf{S}=\begin{pmatrix} \hat{S}_{11} & \hat{S}_{12} \\ \hat{S}_{21} & \hat{S}_{22} \end{pmatrix}=
\begin{pmatrix} \hat{T}^\mathrm{f} & \hat{R}^\mathrm{b'} \\ \hat{R}^\mathrm{f} & \hat{T}^\mathrm{b'} \end{pmatrix}.
\end{equation}
For reciprocal systems we have $\hat{S}_{11}=\hat{S}_{22}^\mathrm{T}$. For the backward direction we added a prime to the transmission and reflection matrices to take into account the flip of the coordinate system when looking in negative $z$-direction as detailed in the appendix.
\section{The stacking}
If all individual layers of the stack possess a negligible reflection, the overall transmission can be obtained by simple multiplication of the individual Jones matrices \cite{Jones1941,YehJones1982,MenzelPRAJones}. However, for resonant periodically structured layers, the assumption of negligible reflection or multiple reflections between the layers is unjustified, except for specific cases like balanced Huygens surfaces\cite{IsaHuygens,GrbicReflectionLess}. When taking into account reflection as well, calculating the overall response of a stack containing polarisation-changing layers, e.g. anisotropic or chiral media or low-symmetry MS, is non-trivial such that analytical formulas of reasonable size can be obtained just for the case of two layers. Hence, the aim of the manuscript is to present a general algorithm applicable to any number of layers with arbitrary symmetry, given in terms of a 4x4 S-matrix.

Once we have the S-matrices of the MS under consideration at hand, we can stack them in an arbitrary manner with arbitrary homogeneous spacer layers in between. Therefore, we need to know not only the S-matrices for the MSs but also the S-matrices $\mathbf{S}_{n,d}$ for propagation in homogeneous media characterized by a refractive index $n$ and thickness $d$ and the S-matrix $\mathbf{S}_{n_1,n_2}$ for the transition between two homogeneous media with refractive indices $n_1$ and $n_2$.
The S-matrix $\mathbf{S}_{n,d}$ for the propagation in a homogeneous medium of thickness $d$ with refractive index $n$ and free space wave-number $k_0$ is given by
\begin{equation}
\mathbf{S}_{n,d}=\exp(ik_0nd)\cdot\mathrm{diag}(1,1,1,1).\label{EQ_S_Prop}
\end{equation}
The S-matrix $\mathbf{S}_{n_1,n_2}$ for the interface between 2 homogeneous media with refractive index $n_1$ and $n_2$ (from 1 to 2) is given by
\begin{equation}
\mathbf{S}_{n_1,n_2}=
\begin{pmatrix}
\frac{2n_1}{n_1+n_2} & 0 & \frac{n_1-n_2}{n_1+n_2} & 0 \\
0 & \frac{2n_1}{n_1+n_2} & 0 & \frac{n_1-n_2}{n_1+n_2} \\
-\frac{n_1-n_2}{n_1+n_2} & 0 & \frac{2n_2}{n_1+n_2} & 0\\
0 & -\frac{n_1-n_2}{n_1+n_2} & 0 & \frac{2n_2}{n_1+n_2}\\
\end{pmatrix}\label{EQ_S_Interface}
\end{equation}
according to the Fresnel formulas for the reflection and transmission at an interface at normal incidence\cite{FresnelFormulas}. The formulas can be extended to layers of anisotropic media or chiral media straightforwardly (see appendix).

\begin{figure}[h]
\begin{center}
\includegraphics[width=84mm,angle=0]{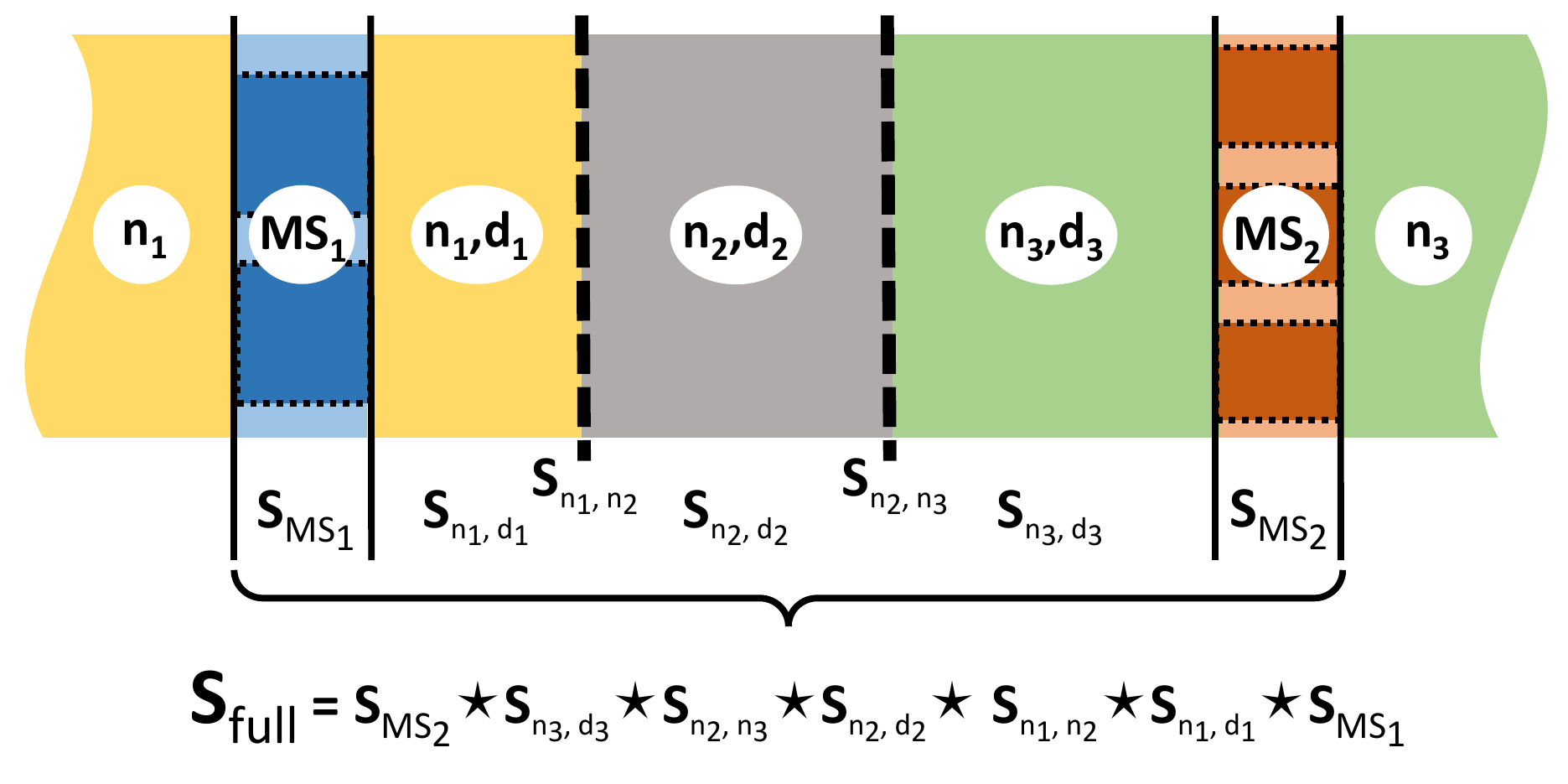}
\caption{Schematic example of two stacked metasurfaces ($\mathrm{MS}_1,\mathrm{MS}_2$) embedded in dielectrics with refractive indices $n_1$ and $n_3$, respectively. Between the metasurfaces there is an additional dielectric layer with refractive index $n_2$ and thickness $d_2$. The general propagation direction is from the front (left) to the back (right). Note the reverse ordering of the S-matrix product. For the overall S-matrix $S_\mathrm{full}$ we have to take into account the S-matrix of $\mathrm{MS}_1$ and $\mathrm{MS}_2$, the propagation through the dielectrics by $S_{n_i,d_i}$ and the transition between the dielectrics by $S_{n_i,n_j}$ at the interfaces marked with bold dashed lines. Note, that the S-matrices for the MS intrinsically contain the transition from and to the dielectric embedding, as their response depends on the embedding via near-field coupling.}
\label{FIG_ExampleSMatrixSystem}
\end{center}
\end{figure}

We can now set up an arbitrary system as shown in Fig.~\ref{FIG_ExampleSMatrixSystem}
by applying the star-product \cite{Li1996} for the connection of S-matrices. For two S-matrices $\mathbf{A}$ and $\mathbf{B}$ it is defined as
\begin{eqnarray*}
\mathbf{A}\star\mathbf{B}&=&\begin{pmatrix} \hat{a}_{11} & \hat{a}_{12} \\ \hat{a}_{21} & \hat{a}_{22} \end{pmatrix}\star\begin{pmatrix} \hat{b}_{11} & \hat{b}_{12} \\ \hat{b}_{21} & \hat{b}_{22} \end{pmatrix}\\
&=&\begin{pmatrix}
\hat{b}_{11}(\mathbb{I}-\hat{a}_{12}\hat{b}_{21})^{-1}\hat{a}_{11} &
\hat{b}_{12}+\hat{b}_{11}\hat{a}_{12}(\mathbb{I}-\hat{b}_{21}\hat{a}_{12})^{-1}\hat{b}_{22}\\
\hat{a}_{21}+\hat{a}_{22}\hat{b}_{21}(\mathbb{I}-\hat{a}_{12}\hat{b}_{21})^{-1}\hat{a}_{11}&
\hat{a}_{22}(\mathbb{I}-\hat{b}_{21}\hat{a}_{12})^{-1}\hat{b}_{22}
\end{pmatrix}
\end{eqnarray*}
By subsequent star-product multiplication we can calculate the S-matrix of an arbitrarily stacked system (see Fig.~\ref{FIG_ExampleSMatrixSystem}).
\subsection{Estimation of the FMA validity}
To guarantee the validity of the FMA is a subtle issue and has to be verified for each MS individually in general. The contribution of the evanescent waves to the reflected and the transmitted field at the distance $d$ to the MS has to be negligible, where e.g. the $x$-polarized field in the transmission at the distance $d$ in Rayleigh expansion \cite{Rayleigh1907} has the form:
$$E_T^x(x,y,d)=\sum_{mn}t^{mn}_{xx}\exp\left[i\left(\frac{2\pi m}{\Lambda_x}x+\frac{2\pi n}{\Lambda_y}y\right)\right]\exp\left[ik_z^{mn}d\right]$$
with the complex transmitted amplitudes $t^{mn}_{xx}$, the propagation constant
$$k_z^{mn}=\sqrt{k_0^2n^2-\left(\frac{2\pi m}{\Lambda_x}\right)^2-\left(\frac{2\pi n}{\Lambda_y}\right)^2}$$
and a refractive index $n$ of the medium in the transmitted region. For simplicity we assumed $t^{mn}_{yx}=0$. Due to the rapid decay of the contribution of the evanescent waves at $z=d$ and the general decay of the amplitudes $t^{mn}_{xx}$ with increasing order $(m,n)$, we can certainly restrict to the consideration of the first evanescent order only, let's say $m=1,n=0$. We approximate the amplitude by $|t^{10}_{xx}|\approx 1$, which is usually valid for MS employing localized resonances. Note that, e.g. for high-Q dielectric waveguide resonances, the amplitude might easily exceed 1, due to the strong field enhancement inside the waveguide. By requiring the  modulus of the evanescent first diffraction order at $z=d$ to be smaller than $e^{-2\pi}\approx 1.8e-3$ we get:
$$e^{-\Im(k_z)d}\le\,e^{-2\pi}\,\rightarrow\,\frac{d}{\lambda}\left[\left(\frac{\lambda}{\Lambda}\right)^2-n^2\right]^{\frac{1}{2}}\ge\,1$$
For a distance $d$ larger than a critical thickness $d_\mathrm{crit}$ defined by the inequality above, we can expect the FMA to be valid. Upon rewriting
\begin{equation}
d_\mathrm{crit}=\Lambda\left/\sqrt{1-\frac{\Lambda^2n^2}{\lambda^2}}\right.\label{EQ_DCrit}
\end{equation}
we see that the critical thickness is diverging at the occurrence of the first diffraction order with $\lambda=n\Lambda$ and approaches $\Lambda$ for $\lambda\gg n\Lambda$, hence monotonically decreasing for increasing $\lambda$. Of course, for systems comprised of MS with different periods and different embedding dielectrics the critical thickness is given by the largest period $\Lambda$, the largest refractive index $n$ and the smallest wavelength $\lambda$.
\subsection{Symmetry operations on S-matrices}
\begin{figure}[h]
\begin{center}
\includegraphics[width=84mm,angle=0]{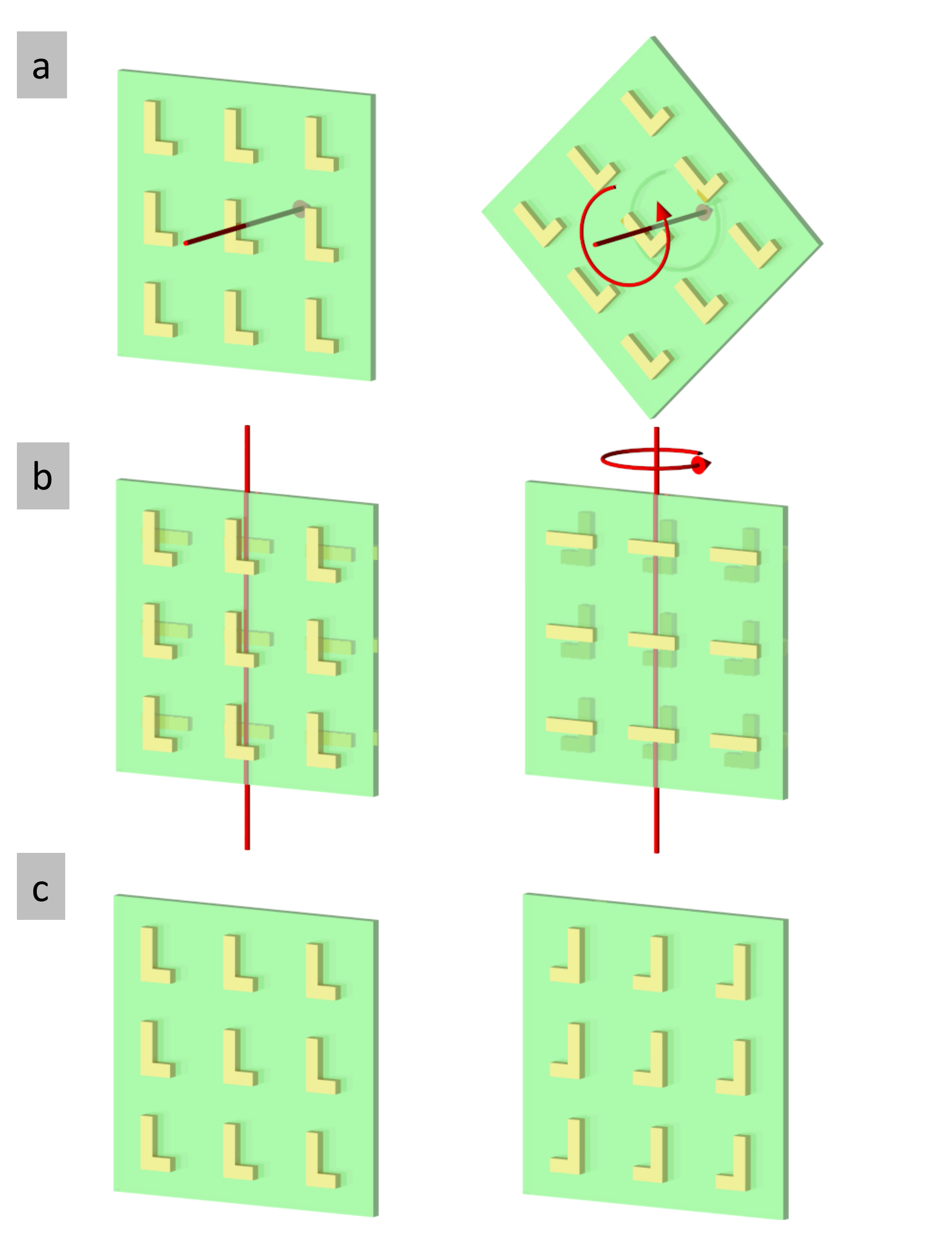}
\caption{Schematic examples for periodically structured metasurfaces. Left column: original system. Right: rotated (a), flipped (b) and mirrored (c) system.}
\label{FIG_Operations}
\end{center}
\end{figure}

Once we have the S-matrix for a specific system, we can analytically calculate the S-matrix for the system when rotated by an angle $\varphi$ around the $z$-axis, or when flipped, i.e. operated from the backside or when mirrored (see. Fig.~\ref{FIG_Operations}). In the following we present the respective expressions.

For an arbitrary matrix $\hat{A}$ the reflection along $x$- or $y$-direction with the respective matrices
\begin{equation}
\hat{M}_x=\begin{pmatrix} -1 & 0 \\ 0 & 1\end{pmatrix}=\hat{M}_x^T,~\hat{M}_y=\begin{pmatrix} 1 & 0 \\ 0 & -1\end{pmatrix}=\hat{M}_y^T
\end{equation}
leads to
\begin{equation}
\hat{M}_x\hat{A}\hat{M}_x=\hat{M}_y\hat{A}\hat{M}_y=\hat{M}\hat{A}\hat{M}
\end{equation}
where $\hat{M}$ is either $\hat{M}_x$ or $\hat{M}_y$.
Mirroring the structure at the $xz$- or $yz$- plane leads to the S-matrix $\mathbf{S}^M$
\begin{equation}
\mathbf{S}^M=\begin{pmatrix}
\hat{M}\hat{S}_{11}\hat{M} & \hat{M}\hat{S}_{12}\hat{M} \\
\hat{M}\hat{S}_{21}\hat{M} & \hat{M}\hat{S}_{22}\hat{M}
\end{pmatrix}.
\end{equation}
Rotating the structure by an arbitrary angle $\varphi$ around the $z$-axis by the rotation matrix
\begin{equation}
\hat{R}_\varphi=\begin{pmatrix} \cos\varphi & \sin\varphi \\ -\sin\varphi & \cos\varphi \end{pmatrix},~\hat{R}_\varphi^T=\hat{R}_{-\varphi}
\end{equation}
leads to $\mathbf{S}^R$
\begin{equation}
\mathbf{S}^R=\begin{pmatrix}
\hat{R}^T\hat{S}_{11}\hat{R} & \hat{R}^T\hat{S}_{12}\hat{R} \\
\hat{R}^T\hat{S}_{21}\hat{R} & \hat{R}^T\hat{S}_{22}\hat{R}\label{EQ_Rotation}
\end{pmatrix}.
\end{equation}
Flipping the structure, i.e. looking at it from the backside leads to $\mathbf{S}^F$
\begin{equation}
\mathbf{S}^F=\begin{pmatrix}
\hat{M}\hat{S}_{22}\hat{M} & \hat{M}\hat{S}_{21}\hat{M} \\
\hat{M}\hat{S}_{12}\hat{M} & \hat{M}\hat{S}_{11}\hat{M}
\end{pmatrix}.
\end{equation}
With these operations we have direct access to the S-matrices of mirrored, flipped and rotated systems without the need for a new rigorous determination.
\section{Exemplary metasurface stacks}
In the following section prototypical examples for stacked MSs are discussed. Particular attention is paid to the error of the stacking compared to rigorous solutions for the stacked systems. To quantify the error we introduce the following quantity:
\begin{equation}
\Delta S_{ij}(d)=\max_{\omega}\left\{|S_{ij}^\mathrm{rig}(\omega,d)|^2-|S_{ij}^\mathrm{stack}(\omega,d)|^2\right\}\label{EQ_Error}
\end{equation}
providing a measure for the deviation between the the rigorous ($S_{ij}^\mathrm{rig}$) and the approximated ($S_{ij}^\mathrm{stack}$) solution of the overall S-matrix within a specific frequency range, which is $100-500$\,THz for all the examples studied here. Hence, the smallest wavelength is $600$\,nm.
\subsection{Stacks of wires}
\begin{figure}[h]
\begin{center}
\includegraphics[width=84mm,angle=0] {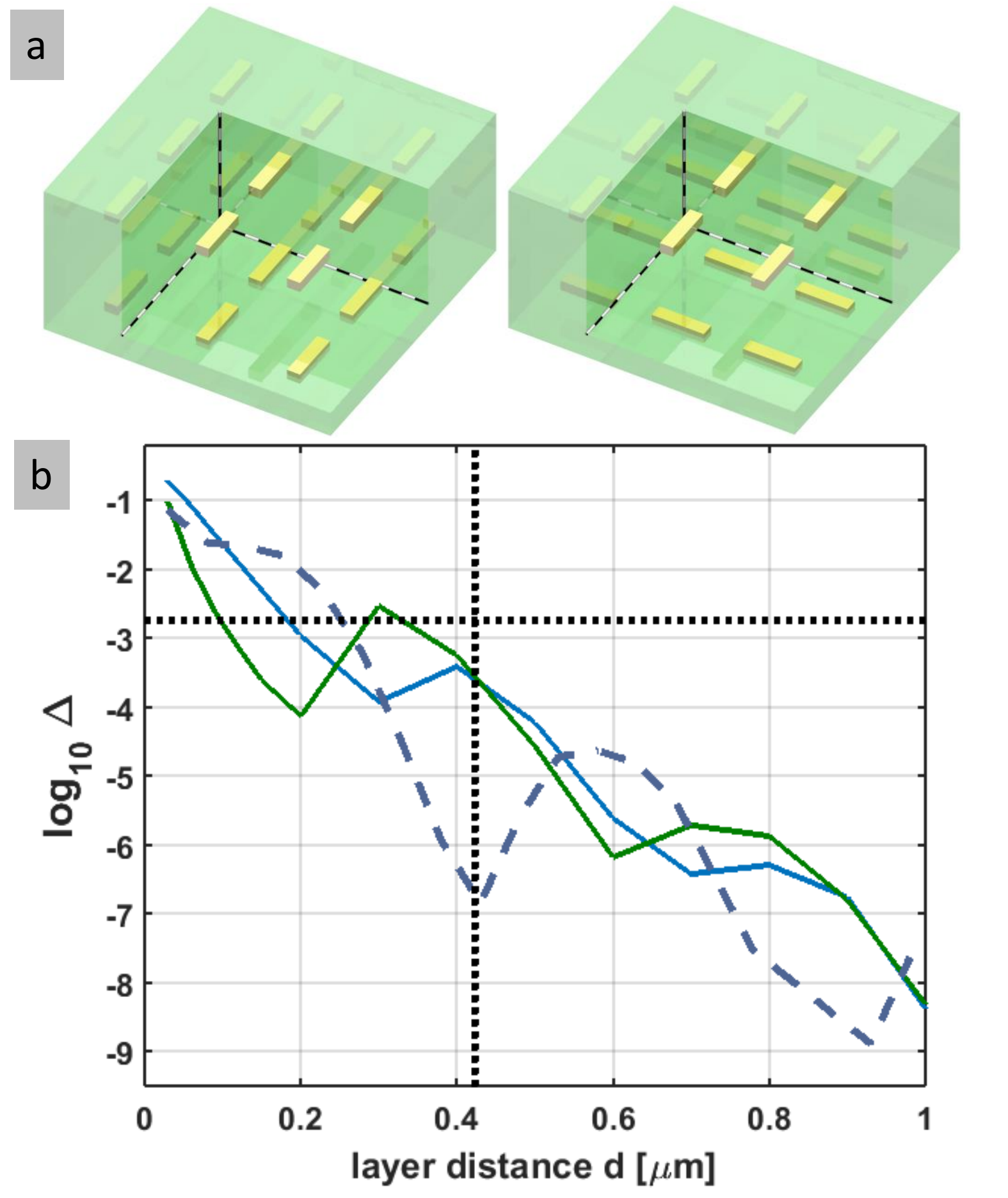}
\caption{a) Schematic of the geometry of parallel (left) and orthogonal (right) wire stacks. b) Decadic logarithm of the maximum error $\Delta S_{ij}(d)$. The solid lines correspond to $\Delta S_{ij}$ for $t_{xx}$ (blue) and $t_{yy}$ (green) of the parallel wire stack. The dashed line corresponds to $\Delta S_{ij}$ for $t_{xx}=t_{yy}$ of the orthogonal wire stack.}
\label{FIG_Bars}
\end{center}
\end{figure}

At first we consider a periodic square array ($\Lambda_x=\Lambda_y=300$\,nm) of resonant plasmonic wires. The wires are made of gold \cite{Johnson1972}, symmetrically embedded in a homogeneous dielectric with $n=1.41$ with a length of $l=240$\,nm, a width of $w=60$\,nm and a height of $h=30$\,nm. The distance between the wire planes in z-direction is varied between $d=30...1000$\,nm. Two different scenarios are investigated with wires oriented parallel and orthogonal to each other (see Fig.\,\ref{FIG_Bars}a).
The S-matrices for the individual MS as well as for the rigorous solution of the stacked system are calculated by FMM \cite{Li1997} directly. For the stacking-algorithm the S-matrix for the symmetrically embedded wires is obtained just once for wires oriented parallel to the $x$-axis. The S-matrix for the $y$-oriented wires are obtained by applying the rotation by $\pi/2$ given in eq.~(\ref{EQ_Rotation}). Together with the S-matrix for the propagation over the distance $d$ in a medium with refractive index $n=1.41$ as given by eq.~(\ref{EQ_S_Prop}), we get the overall S-matrix $S_{ij}^\mathrm{stack}$. The results for the decadic logarithm of the maximum error $\Delta S_{ij}(d)$ are shown in Fig.~\ref{FIG_Bars}b. The solid lines correspond to the transmission for $x$- and $y$-polarized light ($t_{xx},t_{yy}$) of the parallel wires. The dashed line corresponds to the transmission for $x$- and $y$-polarized light for the orthogonal wires, which is the same due to symmetry reasons. Furthermore, for the off-diagonal elements we have $t_{xy}=t_{yx}=r_{xy}=r_{yx}=0$. Since the transmission and the reflection behave similarly with respect to the error, the error in transmission is plotted only. We clearly observe the exponential decay of the error as discussed while deriving the critical thickness $d_\mathrm{crit}$ [see eq.~(\ref{EQ_DCrit})]. We also plotted the critical thickness which is
\begin{equation}
d_\mathrm{crit}=300\,\mathrm{nm}\left/\sqrt{1-\frac{1.41^2\cdot 300^2\,\mathrm{nm}^2}{600^2\,\mathrm{nm}^2}}\right.=423\,\mathrm{nm}\label{EQ_DCrit_Explicit}
\end{equation}
and the limiting error $1.8\mathrm{E}-3$ as black dashed lines. Obviously the estimated critical thickness provides a reasonable measure for the deviation between rigorous and approximated solution. The non-monotonic decrease of the error is due to Fabry-Perot oscillations occurring between the MSs.
\subsection{Stacks of L-shaped particles}
\begin{figure}[h]
\begin{center}
\includegraphics[width=84mm,angle=0] {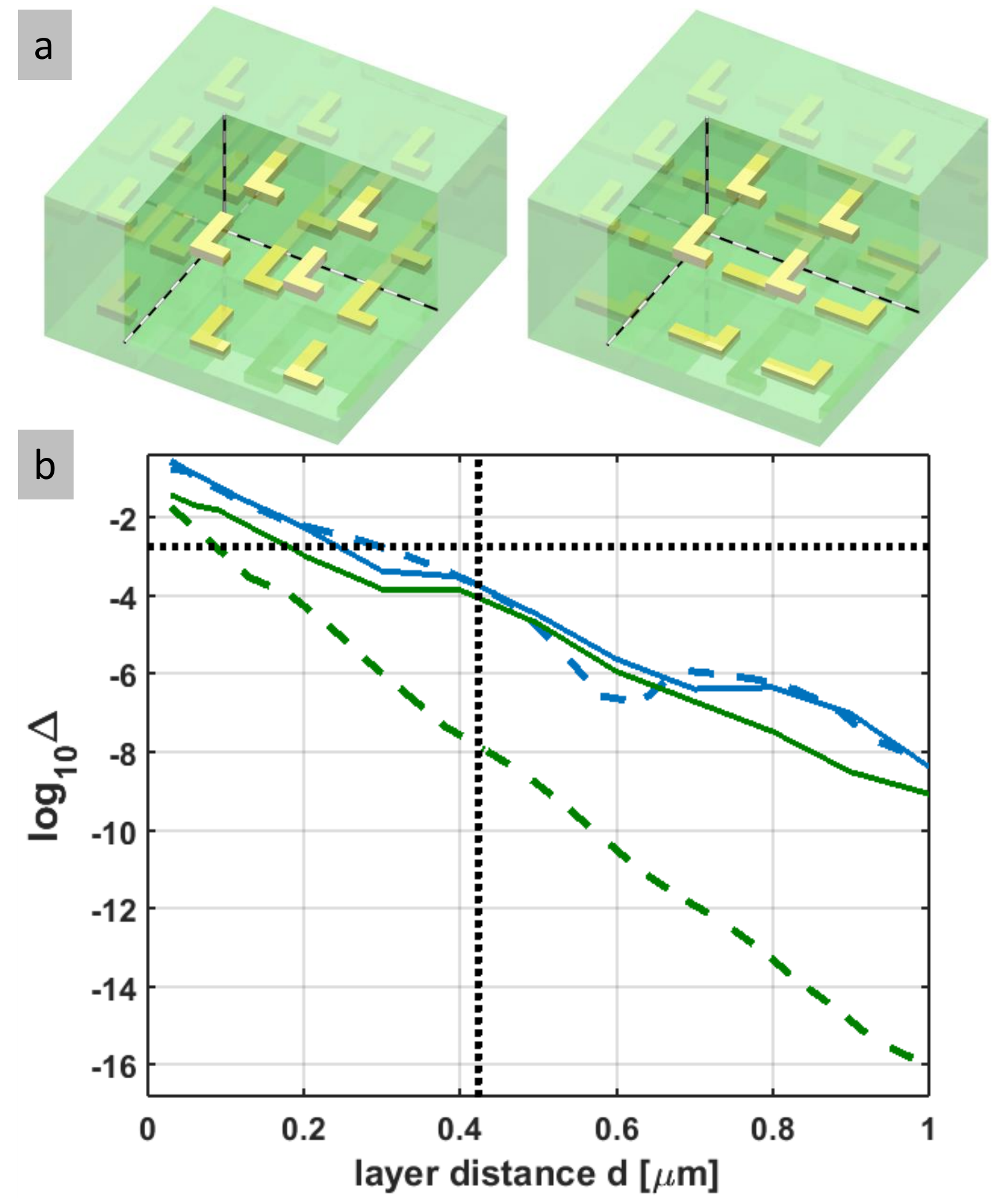}
\caption{a) Schematic of the geometry of parallel (left) and orthogonal (right) L-particle stacks. b) Decadic logarithm of the maximum error $\Delta S_{ij}(d)$. The solid lines correspond to $\Delta S_{ij}$ for $t_{xx}$ (blue) and $t_{xy}$ (green) of the parallel L stack. The dashed lines corresponds to $\Delta S_{ij}$ for $t_{xx}$ (blue) and $t_{xy}$ (green) of the orthogonal L stack.}
\label{FIG_Ls}
\end{center}
\end{figure}

In fact, for the calculation of the overall transmission and reflection for the stacked wires textbook Airy-formulas might have been used due to the non-occurrence of cross-polarized field components. The actual strength of the proposed S-matrix stacking lies in its possibilities for calculating the response of stacked systems exhibiting cross-polarizations, which cannot be handled conveniently by means of analytical formulas. Hence, in a second example we treat the more complex case of stacked resonant plasmonic L-shaped particles (see Fig.~\ref{FIG_Ls}a), which are prototypical metaatoms for polarization control \cite{Lparticle1,Lparticle2}. The asymmetric L's are made of gold, arranged on square lattices with a period of $\Lambda=300$\,nm, with arms length of $240$\,nm and $160$\,nm, a width of $w=60$\,nm and a height of $30$\,nm. They are symmetrically embedded in a dielectric with $n=1.41$. The distance between the layers is variable between $d=30...1000$\,nm. Again, we use parallel and orthogonal oriented L-shaped particle arrays. Again, the S-matrix for the L's is obtained only once. The S-matrix for the rotated L's is obtained by using eq.~(\ref{EQ_Rotation}).
In Fig.~\ref{FIG_Ls}b we have plotted the maximum error between the approximated and the rigorous solution according to eq.~(\ref{EQ_Error}). The solid lines correspond to the transmission $t_{xx}$ and $t_{xy}$ for the parallel L's, the dashed lines correspond to $t_{xx}$ and $t_{xy}$ for the orthogonal L's. Again, the horizontal and the vertical dashed black lines indicate an error of $1.8\mathrm{E}-3$ and the critical thickness of $d_\mathrm{crit}=423$\,nm, respectively. Clearly, the estimated critical thickness gives a reasonable measure for the minimum distance of the layers.
\begin{figure}[h]
\begin{center}
\includegraphics[width=84mm,angle=0] {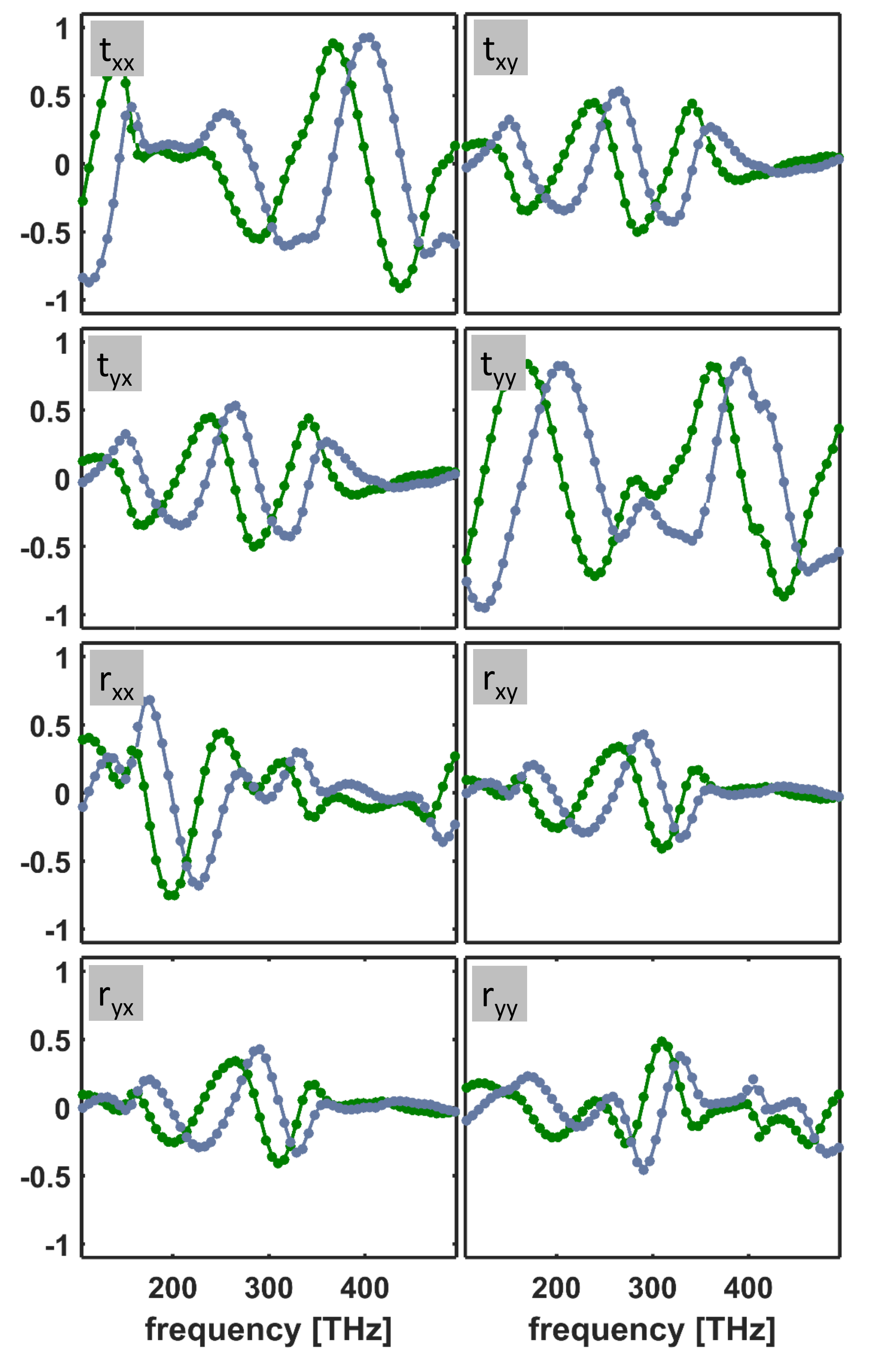}
\caption{Comparison of the rigorous (solid lines) and the approximated (dotted lines) solution for the first two columns of the S-matrix for the stack of parallel L's with as distance of $d=150$\,nm. The graphs show the real (blue) and imaginary (green) parts of the respective co- and cross-polarized reflection and transmission coefficients upon plane wave illumination propagating in +$z$-direction.}
\label{FIG_L00_FMM2SASA}
\end{center}
\end{figure}
For all the S-matrix entries, i.e. co- and cross-polarized transmission and reflection, the linear decrease of the maximum error with increasing distance $d$ between the layers is similar, except for the cross-polarized transmission for orthogonal L's. Here, the decrease with the distance is twice as fast, as the cross-polarization itself is due to the near-field coupling between the two layers only, quickly disappearing for distances $d\gtrapprox 50$\,nm.

To elucidate the actual error and the symmetry of the S-matrix, the real and imaginary parts of the complex S-matrix elements for the parallel L's are plotted in Fig.~\ref{FIG_L00_FMM2SASA} for forward direction, i.e. the first two columns and hence 8 elements of the 4x4 S-matrix. The solid line corresponds to the rigorous solution. The dotted lines correspond to the approximated solution. With respect to the given scale the solutions coincide perfectly for all frequencies even for the small distance of $d=150$\,nm shown here. Due to symmetry the cross-polarized components in transmission ($t_{xy}=t_{yx}$) and reflection ($r_{xy}=r_{yx}$) are identical (achiral), respectively. Due to the lack of rotational symmetries
%(except $C_{2}$)
the diagonal elements are different, showing a strong anisotropy.

\begin{figure}[h]
\begin{center}
\includegraphics[width=84mm,angle=0] {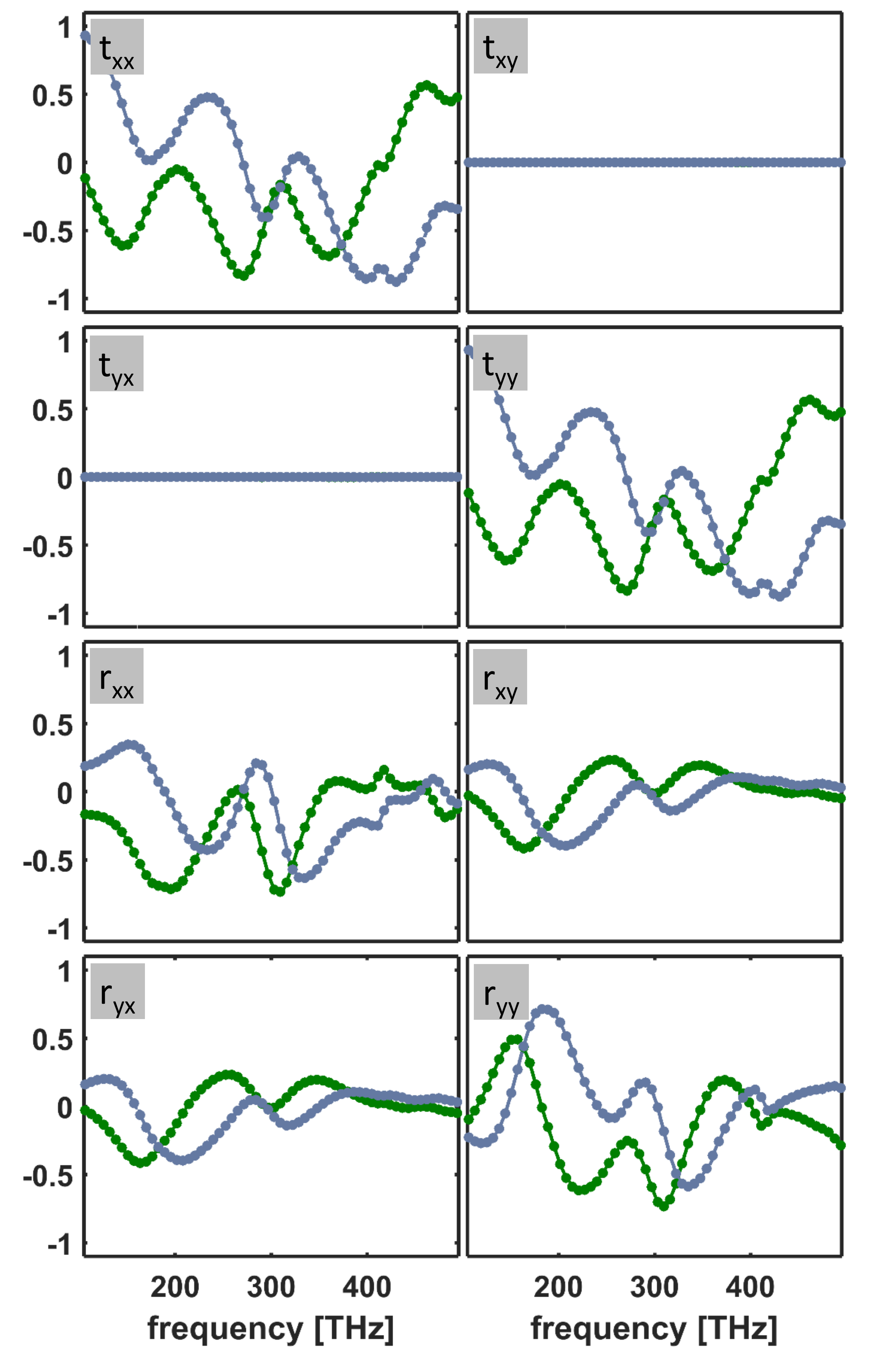}
\caption{Comparison of the rigorous (solid lines) and the approximated (dotted lines) solution for the first two columns of the S-matrix for the stack of orthogonal L's with as distance of $d=150$\,nm. The graphs show the real (blue) and imaginary (green) parts of the respective co- and cross-polarized reflection and transmission coefficients upon plane wave illumination propagating in +$z$-direction.}
\label{FIG_L90_FMM2SASA}
\end{center}
\end{figure}

In Fig.~\ref{FIG_L90_FMM2SASA} we have
plotted the same S-matrix elements for the orthogonal L's. Again the approximated (solid line) and the rigorous solution (dotted line) coincide perfectly at this distance of $d=150$\,nm. The co- and cross-polarized reflection is similar to the case of parallel L's. Quite surprisingly, the transmission shows an unexpected polarization independent behavior with $t_{xx}=t_{yy}$ and $t_{xy}=t_{yx}=0$. The overall structure exhibits no symmetry and is clearly chiral. However, no polarization change occurs as soon as the layers are decoupled with respect to the near-field. Only for distances smaller than $d\lessapprox 50$\,nm a significant polarization occurs as indicated by $t_{xy}$ in Fig.~\ref{FIG_Ls}b. Note, that $t_{xy}=0$ for all distances in the approximated solution.
\subsection{Stacks of particles with different periods}
%
% Wie sind die S-Matrizen bestimmt und wie befuellt. Abstand nur f\"ur 2 Szenarien. Rotation der Schichten. Unterschiedliche Perioden mit grosser \"Uberperiode.
%
One of the major advantages of the stacking formalism is its capability of efficiently treating stacked MSs with different or even incommensurable periods. As a practical example, consider the case of a MS that is supposed to support multiple resonances. One could try to design the individual metaatom such that it supports several resonances. That usually requires the metaatom to be large and eventually not subwavelength anymore. On the other hand, the unit cell could be comprised of several metaatoms in the same layer each addressing a slightly different frequency range. However, such a unit cell would again become to large to be subwavelength. Alternatively, several MS comprised of slightly different metaatoms might be stacked. To keep the density of the metaatoms or the filling fraction in each layer constant, the periods in each layer have to change slightly as well. Unfortunately, their common super-period might get huge and eventually not accessible to rigorous calculations of the overall stack. Here, the stacking algorithm can be used, drastically decreasing the computational efforts.

\begin{figure}[h]
\begin{center}
\includegraphics[width=74mm,angle=0] {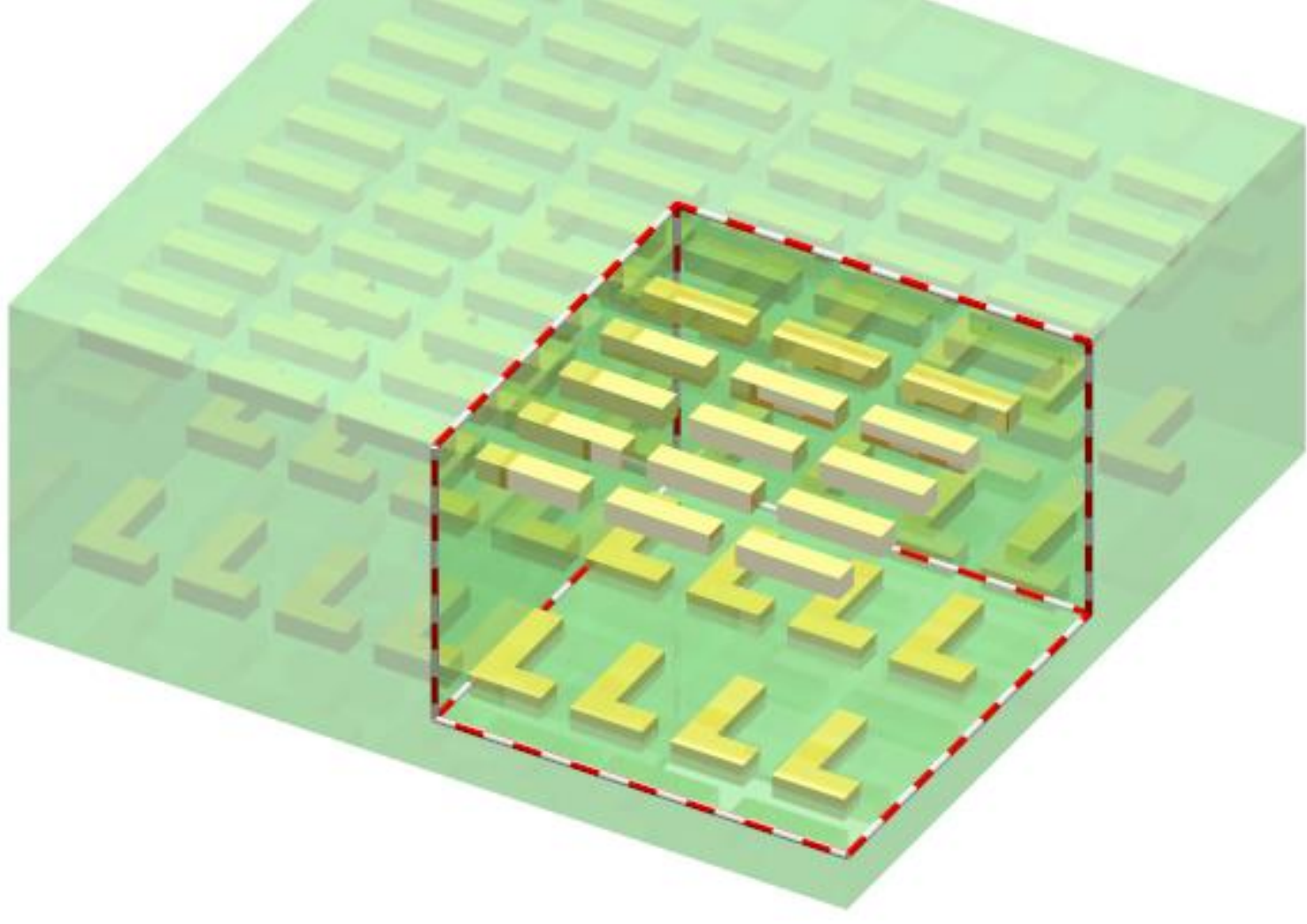}
\caption{Schematic of the geometry of stacked wires and L's. Details of the geometrical parameters are given in the text. Due to the different periods of the wire and the L-particle array, a super-cell (white-red-dashed box) calculation is necessary, containing 5x3 wires and 2x4 L's.}
\label{FIG_GeometryFDTD}
\end{center}
\end{figure}

To give an example and proof the applicability of the method, we consider a stack of arrays of wires and L's (see Fig.~\ref{FIG_GeometryFDTD}) entirely embedded in a dielectric with $n=1.41$. The wires and L's are assumed as gold (see Appendix for the permittivity model).
The array of L's has a period of $\Lambda_x=333.3$\,nm and $\Lambda_y=250$\,nm, arm length along $x$- and $y$-direction of $l_x=250$\,nm and $l_y=180$\,nm, a width of $w=60$\,nm and a height $h=30$\,nm. The array of wires has a period of $\Lambda_x=133.3$\,nm and $\Lambda_y=333.3$\,nm, arm length along $y$-direction of $l_y=200$\,nm, a width of $w=50$\,nm and a height $h=30$\,nm. The arrays have a common super-period of $\Lambda_x=666.6$\,nm and $\Lambda_y=1000$\,nm. This time we use FDTD (MEEP) \cite{MEEP} with a spatial resolution of $2$\,nm for calculating the S-matrices as we need in particular for the super-cell to run FDTD in parallel mode. The S-matrices where built up manually by calculating the $x$- and $y$-polarized zeroth order transmitted $t_{ij}$ the reflected $r_{ij}$ complex fields upon $x$- and $y$-polarized normally incident plane wave excitation. Note that the reflected and transmitted field is defined with respect to planes $20$\,nm in front and behind the structured surfaces, respectively. For the individual arrays of wires and L's a single period was used, drastically decreasing the numerical efforts compared to the super-cell calculation necessary for the stacked system. Furthermore, due to the mirror symmetry with to respect the $xy$-plane and reciprocity of the system the S-matrices for the individual layers were built up based on the transmission and reflection coefficients for illumination in $+z$-direction (forward, first 2 columns) only. For the case of L's we get:
\begin{equation}
\mathbf{S}_\mathrm{L}=
\begin{pmatrix}
\underline{t_{xx}} & \underline{t_{xy}} & r_{xx} & r_{xy} \\
t_{xy} & \underline{t_{yy}} & r_{xy} & r_{yy} \\
\underline{r_{xx}} & \underline{r_{xy}} & t_{xx} & t_{xy} \\
r_{xy} & \underline{r_{yy}} & t_{xy} & t_{yy}
\end{pmatrix},\label{EQ_S_Ls}
\end{equation}
where only the 6 underlined elements had to be determined. The remaining ones are fixed due to reciprocity and mirror symmetry. Furthermore, for the S-matrix of the wires we get
\begin{equation}
\mathbf{S}_\mathrm{wire}=
\begin{pmatrix}
\underline{t_{xx}} & 0 & r_{xx} & 0 \\
0 & \underline{t_{yy}} & 0 & r_{yy} \\
\underline{r_{xx}} & 0 & t_{xx} & 0 \\
0 & \underline{r_{yy}} & 0 & t_{yy}
\end{pmatrix}.\label{EQ_S_Wires}
\end{equation}

The largest period of both arrays is $\Lambda=333.3$\,nm. With the embedding $n=1.41$ and the smallest wavelength of interest of $\lambda=600$\,nm, we find for the critical thickness $d_\mathrm{crit}=536.1$\,nm.
\begin{figure}[h]
\begin{center}
\includegraphics[width=84mm,angle=0] {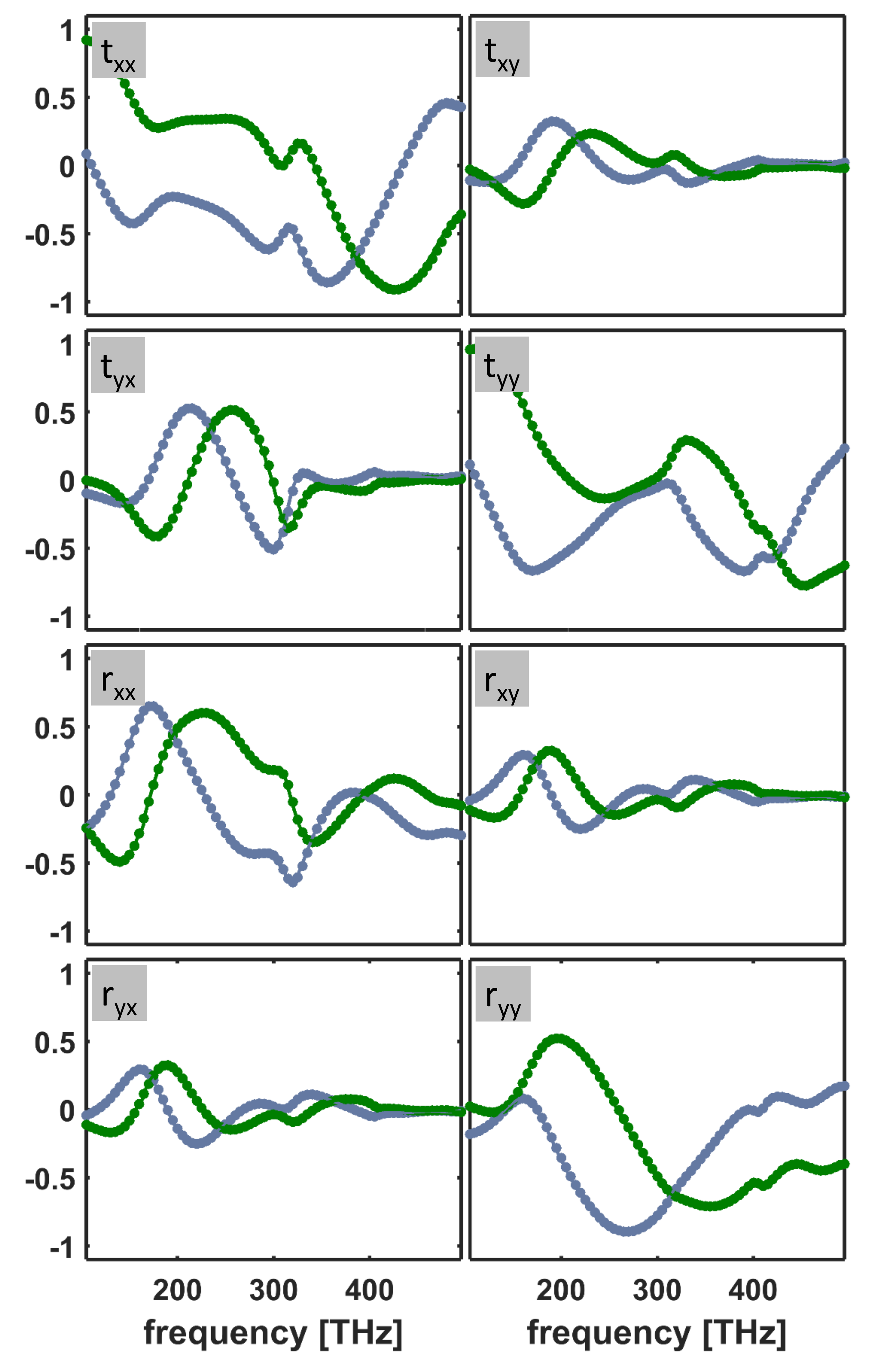}
\caption{Comparison of the rigorous (solid lines) and the approximated (dotted lines) solution for the first two columns of the S-matrix for the stack of L's and wires as shown in Fig.~\ref{FIG_GeometryFDTD} with as distance of $d=250$\,nm. The graphs show the real (blue) and imaginary (green) parts of the respective co- and cross-polarized reflection and transmission coefficients upon plane wave illumination propagating in +$z$-direction.}
\label{FIG_ResultsFDTD}
\end{center}
\end{figure}
As we know from the previous examples, the difference between the rigorous and the approximated solution is sufficiently small already for distance of approx. $d_\mathrm{crit}/2$. Hence, we compared both solutions for a distance between both MS of $d=250$\,nm. The results for the real and the imaginary parts of the forward-part of the S-matrix are shown in Fig.~\ref{FIG_ResultsFDTD}. With respect to the accessible scale the approximated and the rigorous solution are in perfect agreement. In fact the maximum error $\Delta S_{ij}(d=250\,\mathrm{nm})$ is smaller than $0.06$.
\section{Conclusion}
To eventually establish metamaterials as building blocks for modern photonic devices, the optical properties of the individual blocks (metamaterial or metasurface layers) need to be unique, independent of the neighboring ones or their environment. Irrespective of the parameter sets used to describe their optical properties - material parameters, wave parameters or simply their transmission and reflection coefficients concatenated in an S-matrix - their uniqueness requires the MM layers to be homogeneous, i.e. decoupled with respect to the near-field interaction. Otherwise, the optical far-field response of a stack of MM layers cannot be predicted by the far-field response of the individual layers.

To circumvent the introduction and eventually the retrieval of effective parameters in particular for the subtle case of low-symmetry MM layers, we propose here to use the frequency dependent 4x4 S-matrix of the MM layers to fully describe their far-field response upon normally incident plane wave excitation. The far-field response of arbitrary MM stacks can then be determined by use of the adapted S-matrix formalism presented in this contribution. We discussed the range of its applicability, presented a measure for the limits of validity and supported our findings by several examples. We provided all the necessary ingredients for efficiently calculating the response of stacked homogeneous metamaterials and metasurfaces.

The proposed formalism can be applied to any material system, arbitrarily shaped metaatoms, at any frequency and with arbitrary subwavelength periods which can be mutually different as well as incommensurable. In particular in the latter case a rigorous numerical treatment is impossible and the proposed S-matrix formalism is the ultimate choice for calculating the optical far-field response. Combining structured metasurfaces with each other as well as with isotropic, anisotropic or chiral homogeneous layers is possible by simple semi-analytical S-matrix multiplication. Hence, complex stacks and resonators can be set up, accurately treated and optimized with respect to their dispersive polarization sensitive optical functionality without the need for further rigorous full-wave simulations. In that sense, the presented approach is the essence of what is actually possible with homogeneous MS and what MM were designed for.

The proposed stacking formalism can be used for fast and efficient optimization of the optical response of stacked homogeneous MM with respect to a specific dispersion as well as polarization. Complemented by analytical calculations of the S-matrices of the individual layers, we believe that the presented method will open the fast lane towards complex MM engineering.

\section{Acknowledgement}
C.M. gratefully acknowledges support by the Carl-Zeiss foundation. J.S. gratefully acknowledges support the German Federal Ministry of Education and Research (3d sensation). We thank Kay Dietrich for providing the ellipsometric measurements of the gold permittivity and the respective fitting of the permittivity model, Wilm Schumacher for working on the first version of the stacking formalism and Thomas Fl\"ugel-Paul for fruitful discussions and providing the FMM implementation used here.
\section{Appendix}
\subsection{The subtle issue with the coordinate system}
Plainly speaking, the S-matrix contains the complex reflection and transmission coefficients or matrices respectively in forward and backward direction. That's certainly true, however, just for a fixed laboratory coordinate system. The actual 2x2 transmission and reflection matrices $\hat{T}^\mathrm{b}$ and $\hat{R}^\mathrm{b}$ obtained when illuminating the structure from the backside, i.e. within a flipped coordinate system [see Fig.~\ref{FIG_Operations}] are different to the entries of the S-matrix. Rotating the structure around the $x$- or $y$-axis by $180^\circ$ to look at it from the backside leads to a change from $x\rightarrow -x$ or $y\rightarrow -y$. This operation is implemented by the reflection matrix
$$\hat{M}=\begin{pmatrix} -1 & 0 \\ 0 & 1 \end{pmatrix}=\hat{M}^T,$$
given here for the rotation around the $x$-axis. Note, that the rotation around the $y$-axis gives identical results, as the subsequent rotation around $z$ by $180^\circ$ does not affect the S-matrix.
Hence, the actual transmission and reflection matrices are
\begin{eqnarray}\hat{T}^\mathrm{b}&=&\begin{pmatrix} t^\mathrm{b}_{xx} & t^\mathrm{b}_{xy} \\ t^\mathrm{b}_{yx} & t^\mathrm{b}_{yy}\end{pmatrix}=\begin{pmatrix} t^\mathrm{b'}_{xx} & -t^\mathrm{b'}_{xy} \\ -t^\mathrm{b'}_{yx} & t^\mathrm{b'}_{yy}\end{pmatrix}=\hat{M}\hat{T}^\mathrm{b'}\hat{M}=\hat{M}\hat{S}_{22}\hat{M}\\
\hat{R}^\mathrm{b}&=&\begin{pmatrix} r^\mathrm{b}_{xx} & r^\mathrm{b}_{xy} \\ r^\mathrm{b}_{yx} & r^\mathrm{b}_{yy}\end{pmatrix}=\begin{pmatrix} r^\mathrm{b'}_{xx} & -r^\mathrm{b'}_{xy} \\ -r^\mathrm{b'}_{yx} & r^\mathrm{b'}_{yy}\end{pmatrix}=\hat{M}\hat{R}^\mathrm{b'}\hat{M}=\hat{M}\hat{S}_{12}\hat{M}.
\end{eqnarray}
Let's consider the S-matrix for the system rotated by an angle $\varphi$ around the propagation direction. Intuitively the rotation from the backside is accomplished by rotation with $-\varphi$. By using the rotation matrix
$$\hat{R}_\varphi=\begin{pmatrix} \cos\varphi & \sin\varphi \\ -\sin\varphi & \cos\varphi \end{pmatrix},~\hat{R}_\varphi^T=\hat{R}_{-\varphi}$$
we get for the front direction:
\begin{equation}
\hat{T}^{\mathrm{f},\varphi}=\hat{R}_{-\varphi}\hat{T}^\mathrm{f}\hat{R}_{\varphi},~\hat{R}^{\mathrm{f},\varphi}=\hat{R}_{-\varphi}\hat{R}^\mathrm{f}\hat{R}_{\varphi}.
\end{equation}
For the backward direction we get
\begin{eqnarray}
\left(\hat{T}^{\mathrm{b},\varphi}\right)'&=&\hat{M}\hat{T}^{\mathrm{b},\varphi}\hat{M}\\
&=&\hat{M}\hat{R}_{\varphi}\hat{T}^\mathrm{b}\hat{R}_{-\varphi}\hat{M}\\
&=&\hat{M}\hat{R}_{\varphi}\hat{M}\hat{T}^\mathrm{b'}\hat{M}\hat{R}_{-\varphi}\hat{M}\\
&=&\hat{R}_{-\varphi}\hat{T}^\mathrm{b'}\hat{R}_{\varphi}
\end{eqnarray}
and
\begin{equation}
\left(\hat{R}^{\mathrm{b},\varphi}\right)'=\hat{R}_{-\varphi}\hat{R}^\mathrm{b'}\hat{R}_{\varphi}.
\end{equation}
Hence, the rotation of the backward matrices is done precisely as for the forward matrices. The intuitive rotation with negative rotation angle is accounted for by the flip of the coordinate system. Note that the subsequent reflection along $x$ and $y$ or vice versa is identical to a rotation by $\varphi=\pi$ and has no effect on the S-matrix.\\
If we introduce the matrices containing the reflection and transmission matrices as obtained in the physically intuitive system of looking in forward and backward direction we get
\begin{equation}
\begin{pmatrix}
\hat{T}^\mathrm{f} & \hat{R}^\mathrm{b} \\
\hat{R}^\mathrm{f} & \hat{T}^\mathrm{b}
\end{pmatrix}=
\begin{pmatrix}
\hat{S}_{11} & \hat{M}\hat{S}_{12}\hat{M} \\
\hat{S}_{21} & \hat{M}\hat{S}_{22}\hat{M}
\end{pmatrix}.
\end{equation}
For the flipped system we get:
\begin{equation}
\begin{pmatrix}
\hat{T}^\mathrm{f} & \hat{R}^\mathrm{b} \\
\hat{R}^\mathrm{f} & \hat{T}^\mathrm{b}
\end{pmatrix}=
\begin{pmatrix}
\hat{M}\hat{S}_{22}\hat{M} & \hat{S}_{21} \\
\hat{M}\hat{S}_{12}\hat{M} & \hat{S}_{11}
\end{pmatrix}
\end{equation}
in accordance with the physical intuition of a simple exchange of $f$- and $b$-matrices.
\subsection{S-matrices for anisotropic and chiral layers}
The S-matrix for propagation over distance $d$ in an anisotropic medium, whose crystal axes are coinciding with the principal coordinate system and with refractive index pair $\mathbf{n}=(n_x,n_y)$ for propagation along $z$-direction, is given by
\begin{equation}
\mathbf{S}_{\mathbf{n},d}=
\begin{pmatrix}
P_x & 0 & 0 & 0 \\
0 & P_y & 0 & 0 \\
0 & 0 & P_x & 0 \\
0 & 0 & 0 & P_y
\end{pmatrix}\label{EQ_S_PropAniso}
\end{equation}
with the propagator $P_i=\exp[ik_0n_id]$.\\
If the crystal is rotated around $z$ with respect to the principal coordinate system, the corresponding S-matrix can be obtained by using eq.~(\ref{EQ_Rotation}).

The S-matrix for the interface between two anisotropic layers (1,2) with the same crystal axes aligned to the principal coordinate system and refractive index pairs $\mathbf{n}_i=(n_{xi},n_{yi})$, is given by
\begin{equation}
\mathbf{S}_{\mathbf{n}_{1},\mathbf{n}_{2}}=
\begin{pmatrix}
\frac{2n_{x1}}{n_{x1}+n_{x,2}} & 0 & \frac{n_{x1}-n_{x2}}{n_{x1}+n_{x2}} & 0 \\
0 & \frac{2n_{y1}}{n_{y1}+n_{y2}} & 0 & \frac{n_{y1}-n_{y2}}{n_{y1}+n_{y2}} \\
-\frac{n_{x1}-n_{x2}}{n_{x1}+n_{x2}} & 0 & \frac{2n_{x2}}{n_{x1}+n_{x2}} & 0\\
0 & -\frac{n_{y1}-n_{y2}}{n_{y1}+n_{y2}} & 0 & \frac{2n_{y2}}{n_{y1}+n_{y2}}\\
\end{pmatrix}\label{EQ_S_InterfaceAniso}
\end{equation}

The more sophisticated case of anisotropic layers ($\mathbf{n}_1,\mathbf{n}_2$) with crystal axes rotated by an angle $\varphi_1$ and $\varphi_2$ [see eq.~(\ref{EQ_Rotation})] can be obtained by taking the star-product of the rotated interface S-matrices between an arbitrary isotropic medium with $n_0$ and the anisotropic medium $\mathbf{n}_i$:
\begin{equation}
\mathbf{S}^{\varphi_2}_{n_0,\mathbf{n}_2}\star\mathbf{S}^{\varphi_1}_{\mathbf{n}_1,n_0}.
\end{equation}

If bi-isotropic chiral layers with refractive index $n$ and chirality parameter $\kappa$ are used, the following S-matrix for the propagation has to be used
\begin{equation}
\mathbf{S}_{n,\kappa,d}=
\exp[ik_0nd]\begin{pmatrix}
\cos\varphi & \sin\varphi & 0 & 0 \\
-\sin\varphi & \cos\varphi & 0 & 0 \\
0 & 0 & \cos\varphi & -\sin\varphi \\
0 & 0 & \sin\varphi & \cos\varphi
\end{pmatrix}\label{EQ_S_PropChiral}
\end{equation}
with $\varphi=k_0\kappa d$. For the interface from and to chiral media the standard isotropic interface S-matrix of eq.~(\ref{EQ_S_Interface}) can be used.
\subsection{Permittivity of gold used for FDTD}
For the FDTD calculations performed with short pulse excitation we had to model the permittivity by a Drude and a Lorentzian term as
\begin{equation}
\varepsilon(\omega)=\varepsilon_\infty+\frac{\delta_1}{-\omega^2-i\gamma_1\omega}+\frac{\delta_2}{-\omega^2-i\gamma_2\omega+c_2}
\end{equation}
with $\omega=2\pi/\lambda$ in $[\mu m^{-1}]$. The normalized parameters are $\varepsilon_\infty=5.53$, $\delta_1=2178.43$, $\gamma_1=0.30978$, $\delta_2 = 465.79$, $\gamma_2=2.94869$ and $c_2=228.713$.
\begin{figure}[h]
\begin{center}
\includegraphics[width=65mm,angle=0] {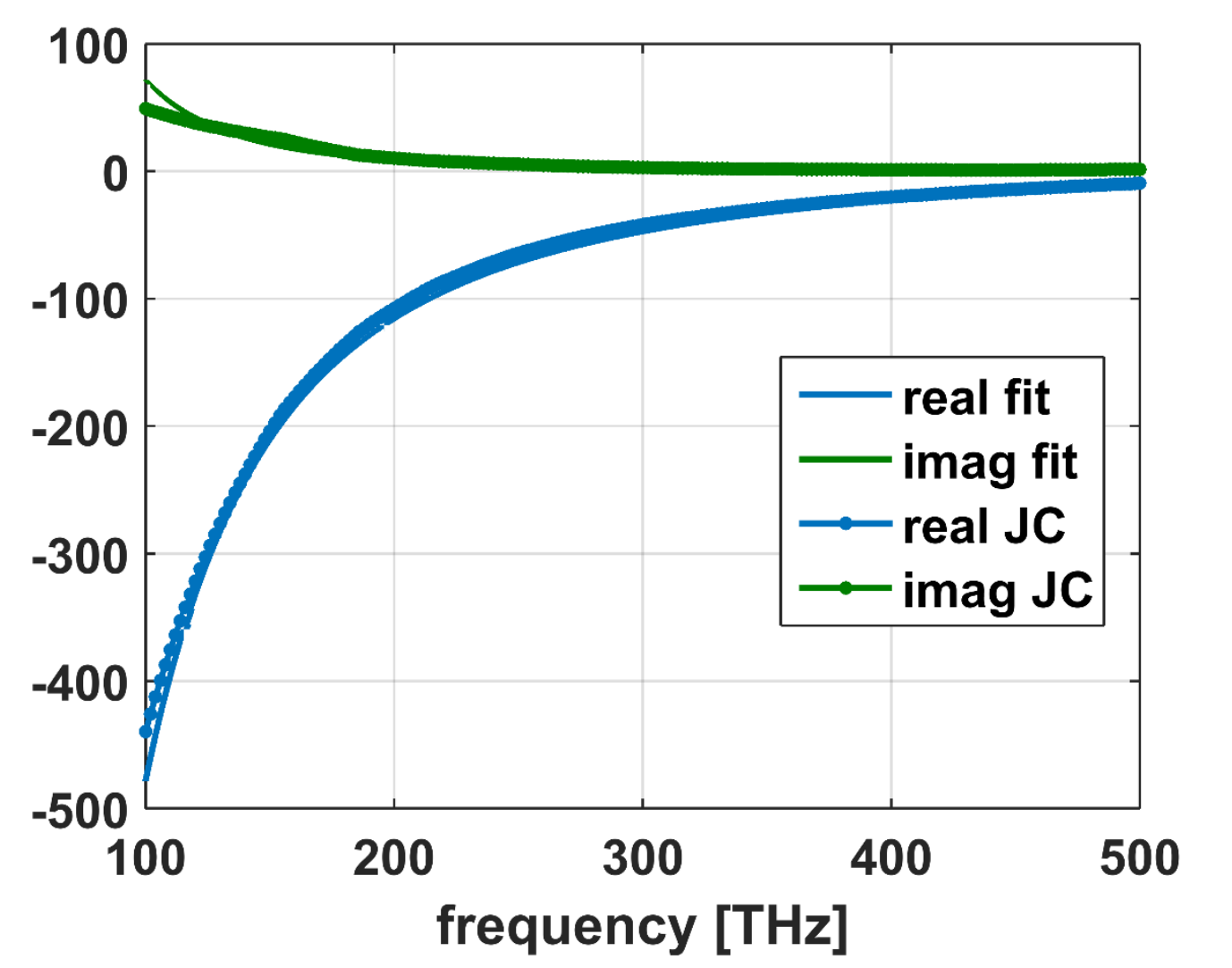}
\caption{Comparison of the real and imaginary part of the permittivity $\varepsilon(\omega)$ as obtained by the ellipsometric fit and the tabulated data from Johnson and Christy \cite{Johnson1972}.}
\label{FIG_CompareJC2FitEpsilon}
\end{center}
\end{figure}
The fit as shown in Fig.~\ref{FIG_CompareJC2FitEpsilon} is performed on ellipsometric data of in-house made gold, in very good agreement with Johnson-Christy data\cite{Johnson1972}. The fit just slightly overestimates the imaginary part of $\varepsilon(\omega)$ close to frequencies around $100$\,THz.

\end{document}